\documentclass[pra,aps,groupaddress,prd]{revtex4}
\usepackage{amsfonts}
\usepackage{epsfig,amsmath,graphicx,amssymb,overpic,bm}
\usepackage{color,tikz,xcolor,extarrows}
\usetikzlibrary{arrows}

\def\be{\begin{equation}}
\def\ee{\end{equation}}
\def\bee{\begin{eqnarray}}
\def\ene{\end{eqnarray}}
\def\bes{\begin{subequations}}
\def\ees{\end{subequations}}

\newcommand{\bR}{{\rm R}}
\newcommand{\mQ}{\mathcal{Q}}
\newcommand{\bQ}{{\bf Q}}
\newcommand{\bPT}{{\mathcal P}{\mathcal T}}
\newcommand{\bP}{{\mathcal P}}
\newcommand{\bT}{{\mathcal T}}

\begin{document}

\title{Integrable $\bPT$-symmetric local and nonlocal vector nonlinear Schr\"odinger equations: a unified two-parameter model}
\author{Zhenya Yan}
\email{zyyan@mmrc.iss.ac.cn}
\affiliation{Key Laboratory of Mathematics Mechanization, Institute of Systems
Science, AMSS, Chinese Academy of Sciences, Beijing 100190, China }

\begin{abstract}
{\bf Abstract}\,\,We introduce a new unified two-parameter $\{(\epsilon_x, \epsilon_t)\,|\epsilon_{x,t}=\pm1\}$  wave model (simply called $\mQ_{\epsilon_x,\epsilon_t}^{(n)}$ model), connecting integrable local and nonlocal vector nonlinear Schr\"odinger equations.
The two-parameter $(\epsilon_x, \epsilon_t)$ family also brings insight into a one-to-one connection  between four points $(\epsilon_x, \epsilon_t)$ (or complex numbers $\epsilon_x+i\epsilon_t$) with $\{{\mathcal I}, \bP, \bT, \bPT\}$ symmetries for the first time. The $\mQ_{\epsilon_x,\epsilon_t}^{(n)}$ model with $(\epsilon_x, \epsilon_t)=(\pm 1, 1)$ is shown to possess a Lax pair and infinite number of conservation laws, and to be $\bPT$ symmetric. Moreover, the Hamiltonians with  self-induced potentials are shown to be $\bPT$ symmetric only for $\mQ_{-1,-1}^{(n)}$ model and to be $\bT$ symmetric only for $\mQ_{+1,-1}^{(n)}$ model. The multi-linear form and some self-similar solutions are also given for the $\mQ_{\epsilon_x,\epsilon_t}^{(n)}$ model including bright and dark solitons, periodic wave solutions, and multi-rogue wave solutions.

\vspace{0.1in}
\noindent {\it Keywords:}  two-parameter family of nonlocal vector nonlinear Schr\"odinger equations; Lax pair; conservation laws; $\bPT$ symmetry; solitons.

\end{abstract}

\maketitle

\baselineskip=12pt


{\it Introduction.}---Some non-Hermitian parity-time ($\bPT$)-symmetric Hamiltonians have been found to possess entirely real spectra and drawn much attention~\cite{pt}. Here the parity reflection operator $\mathcal{P}:$\, $p\rightarrow -p$,\, $x\rightarrow -x$ is linear whereas the time reflection operator $\mathcal{T}:$\, $p\rightarrow -p$,\, $t\rightarrow -t$,\, $i\rightarrow -i$ is anti-linear~\cite{ptv}.
Since $p\rightarrow -p$ is implied in {\it rest} both $\bP$ and $\bT$ symmetries, thus
 it is {\it unnecessary} such that $\bP$ and $\bT$ symmetries can be simplified as $\bP:$ $x\rightarrow -x$ and $\bT:$ $(t,\, i)\rightarrow (-t, -i)=-(t,\, i)$, where we cancel identical maps $(t,\, i)\rightarrow (t,\, i)$ in $\bP$ and $ x\rightarrow x$ in $\bT$. Therefore, we can rewrite a single $\bPT$-symmetric operator $H(x,t)$ as $\bPT H(x,t,i)=H(-x,-t,-i)$. For example, $H(x,t,i)=p^2+U(x,t)$ with $U(x,t)=V(x,t)+iW(x,t)$ and $V(x,t),\, W(x,t)\in \mathbb{R}[x,t]$ is $\bPT$-symmetric, if the necessary
(but not sufficient) condition $V(x,t)=V(-x,-t)$ and $W(x,t)=-W(-x,-t)$ holds. In particular, if the potential $U(x,t)$ only depends on space $x$, i.e., $U(x,t)=U(x)=V(x)+iW(x)$, then a necessary (but not sufficient) condition for $\bPT$-symmetry is that its real and imaginary parts are even and odd functions, respectively. For example, the Bender-Boettcher potential $U(x)=x^2(ix)^{\nu}$~\cite{pt}, the Scarff II potential~\cite{scar} $U(x)=v_0\,{\rm sech}^2(x)+iw_0\,{\rm sech}(x)\tanh(x)$, and the Rosen-Morse potential~\cite{rm} $U(x)=v_0\,{\rm sech}^2(x)+iw_0\tanh(x)$ are all
$\bPT$ symmetric, where $\nu, v_0, w_0$ are real-valued parameters.

Except that some properties related to non-Hermitian $\bPT$-symmetric Hamiltonians with a wide class of potentials have attracted much attention from the theoretical view (see~\cite{Muss, Dis, li, kon, abc, mr}), there have had some experiments to observe $\bPT$-symmetric phase transitions in optical couplers with $\mathrm{Al}_{x}\mathrm{Ga}_{1-x}$~\cite{exp1} and Fe-doped $\mathrm{LiNbO}_{3}$~\cite{exp2}, respectively, microwave billiard~\cite{exp3}, and large-scale temporal lattices~\cite{exp4}, microring resonators~\cite{exp5}, a coherent perfect absorber with $\bPT$ phase transition~\cite{exp7}. But it still an important subject to theoretically explore new $\bPT$-symmetric properties and $\bPT$-symmetric nonlinear waves. At the same time, it is also significant to find integrable $\bPT$-symmetric nonlinear wave models in the study of both $\bPT$-symmetric waves and integrable systems~\cite{ab}.

{\it A unified two-parameter physical model.}---In this Letter, we introduce and investigate in detail a new unified two-parameter $(\epsilon_x,\epsilon_t)$ model (simply called $\mQ_{\epsilon_x,\epsilon_t}^{(n)}$ model)
\vspace{-0.03in} \bee  \label{vnlsp}
i\bQ_t(x,t)=-\bQ_{xx}(x,t)+2\sigma \bQ(x,t)\bQ^{\dag}(\epsilon_xx,\epsilon_tt)\bQ(x,t),\quad
\ene
connecting local and {\it new} nonlocal vector nonlinear Schr\"odinger (NLS) equations, where $\bQ(x,t)=(q_1(x,t), q_2(x,t),\cdots, q_n(x,t))^T$ is a complex-valued column vector, $x, t\in \mathbb{R}$,\, $\epsilon_{x,t}=\pm 1$ are symmetric parameters,\, $\sigma=\pm 1$ denotes the real-valued self-focusing ($-$) and defocusing ($+)$ nonlinear interactions, and $\bQ^{\dag}(\epsilon_x x,\epsilon_tt)$ denotes the transpose conjugate of $\bQ(\epsilon_x x,\epsilon_tt)$.  The $\mathcal{Q}_{\epsilon_x,\epsilon_t}^{(n)}$ model exhibits four distinct waves for two-parameter family $(\epsilon_x, \epsilon_t)\in\{(1,1), (-1,1), (1, -1), (-1, -1)\}$. The $\mQ_{+1,+1}^{(n)}$ model is local and just known integrable vector NLS equations, $\mathcal{Q}_{+1,+1}^{(n)}$ model, including the Manakov system~\cite{vnls, vnls2, vnlsb}. However, $\mQ_{\epsilon_x,\epsilon_t}^{(n)}$ model (\ref{vnlsp}) with $(\epsilon_x, \epsilon_t)\in\{(-1,1), (1, -1), (-1, -1)\}$ is new and nonlocal. We find that the self-induced potentials $\bQ^{\dag}(\epsilon_xx,\epsilon_tt)\bQ(x,t)$ with $(\epsilon_x, \epsilon_t)\in\{(-1,1), (1, -1), (-1, -1)\}$ may {\it not} be real-valued functions, and differ from the real-valued function $\bQ^{\dag}(x,t)\bQ(x,t)$ in $\mathcal{Q}_{+1,+1}^{(n)}$ model. The quasi-power defined by $Q_{\epsilon_x, \epsilon_t}(t)=\int_{-\infty}^{+\infty}\bQ^{\!\dag}(\epsilon_xx,\epsilon_tt)\bQ(x,t)dx$ is conserved during evolution, however the total power of Eq.~(\ref{vnlsp}) defined by $P_{\epsilon_x, \epsilon_t}(t)=\int_{-\infty}^{+\infty}|\bQ(x,t)|^2dx$ is {\it not} conserved during evolution except that the power $P_{+1,+1}(t)$ is conserved during evolution since $dP_{\epsilon_x, \epsilon_t}(t)/(dt)\!=\!-\!2i\sigma$
$\times \!\int_{-\infty}^{+\infty}\!\!dx|\bQ(x,t)|^2\![\bQ^{\!\dag}\!(\epsilon_xx,\epsilon_tt)\bQ(x,t)\!\!-\!\!
\bQ^{\!\dag}(x,\!t)\bQ(\epsilon_xx,\!\epsilon_tt)]$. Therefore, Eq.~(\ref{vnlsp}) differs from the considered NLS equation with $\bPT$-symmetric potential~\cite{Muss}. Eq.~(\ref{vnlsp}) is associated with a variational principle $\delta \mathcal{L}/\delta q_j^{*}(\epsilon_xx,\epsilon_tt)=0\, (j=1,2,...,n)$ with the Lagrangian density
\vspace{-0.03in}
\bee
\begin{array}{l}\mathcal{L}=\!i[\bQ_t^{\!\dag}(\epsilon_xx,\epsilon_tt)\bQ(x,t)-\bQ^{\!\dag}(\epsilon_xx,\epsilon_tt)\bQ_t(x,t)]
  +2\bQ_x^{\!\dag}(\epsilon_xx,\epsilon_tt)\bQ_x(x,t) \vspace{0.1in} \\
 \qquad  +4\sigma\!\!\sum_{i,j=1;\,j\geq i}^nq_i(x,t)q_i^{*}(\epsilon_xx,\epsilon_tt)q_j(x,t)q_j^{*}(\epsilon_xx,\epsilon_tt).
\end{array}\ene

In what follows we investigate the $\mQ_{\epsilon_x,\epsilon_t}^{(n)}$ model with $(\epsilon_x, \epsilon_t)\in\{(-1,1), (1, -1), (-1, -1)\}$ in details.  Eq.~(\ref{vnlsp}) with $n=1$ yields the nonlocal NLS equation ($\mQ_{\epsilon_x,\epsilon_t}^{(1)}$ model)
\bee\label{nls1}
 iq_{1,t}(x,t)\!=\!-q_{1,xx}(x,t)\!+\!2\sigma q_1^2(x,t)q_1^{*}(\epsilon_x x,\epsilon_tt),\quad
\ene
where the star sands for complex conjugate, which with $(\epsilon_x, \epsilon_t)=(-1,1)$ reduces to the known integrable nonlocal NLS equation presented recently in Ref.~\cite{am}. But Eq.~(\ref{nls1}) with $(\epsilon_x, \epsilon_t)\in\{(1, -1), (-1, -1)\}$ are both {\it new}.  Eq.~(\ref{vnlsp}) with $n=2$ yields {\it new} nonlocal vector NLS equations ($\mQ_{\epsilon_x,\epsilon_t}^{(2)}$ model)
\bee\label{nls2} \begin{array}{l}
iq_{j,t}(x,t)\!=\!-q_{j,xx}(x,t)+2\sigma [q_1(x,t)q_1^{*}(\epsilon_xx,\epsilon_tt) +q_2(x,t)q^{*}_2(\epsilon_x x,\epsilon_tt)]q_j(x,t), \, (j=1,2) \qquad
  \end{array}\ene
 In fact, Eq.~(\ref{vnlsp}) with $n>1$  are all {\it new} nonlocal vector NLS equations. In particular, $\mQ_{\epsilon_x,\epsilon_t}^{(2)}$ model with
$(\epsilon_x,\epsilon_t)=(1,1)$ is just the well-known Manakov system~\cite{vnls}.

{\it Lax pair and infinite number of conservation laws.}---We consider the following linear spectral problem~\cite{vnlsb}
\vspace{-0.03in}\bes\label{lax} \bee
\label{lax1}\Psi_x=(-i\lambda \Sigma_3+U)\Psi, \hspace{1.35in} \vspace{0.1in}\\
\label{lax2}\Psi_t=(-2i\lambda^2\Sigma_3+2\lambda U-i U_x\Sigma_3-iU^2\Sigma_3)\Psi,
\ene\ees
where $\Psi=(\psi_1(x,t),\psi_2(x,t),...,\psi_{n+1}(x,t))^T$ is a column eigenvector, $\lambda\in \mathbb{C}$ is a spectral parameter, the generalized Pauli matrix $\Sigma_3$  and potential matrix $U$ are given by
\bee \nonumber
 \Sigma_3=\left(\begin{matrix}  I_n  &  0 \\ 0 & -1 \end{matrix} \right), \,\,\,
 U(x,t)= \left(\begin{matrix}  0_n &  \bQ(x,t) \\ \sigma \bR(x,t) & 0 \end{matrix} \right), \,\,
\ene
where $I_n$ and  $ 0_n$ are $n\times n$ unity and zero matrixes, respectively, and $\bR(x,t)=(r_1(x,t), r_2(x,t),\cdots, r_n(x,t))$ is a complex-valued row vector.  The compatibility condition of Eqs.~(\ref{lax1}) and (\ref{lax2}), $\Psi_{xt}=\Psi_{tx}$, leads to $2n$-component nonlinear wave equations
\vspace{-0.05in}\bes \label{vnls}\bee\label{vnlsa}
i\bQ_t(x,t)=-\bQ_{xx}(x,t)-2\sigma \bQ(x,t)\bR(x,t)\bQ(x,t), \quad\,\, \\
-i\bR_t(x,t)=-\bR_{xx}(x,t)-2\sigma \bR(x,t)\bQ(x,t)\bR(x,t).\quad\,\,
\label{vnlsb}\ene\ees
System (\ref{vnls}) reduces to vector NLS equations for the symmetric reduction $\bR(x,t)=-\bQ^{\dag}(x,t)$~\cite{vnlsb}
\vspace{-0.05in} \bee\label{nls0}
i\bQ_t(x,t)=-\bQ_{xx}(x,t)+2\sigma \bQ(x,t)\bQ^{\dag}(x,t)\bQ(x,t), \quad
\ene
which corresponds to Eq.~(\ref{vnlsp}) with $(\epsilon_x, \epsilon_t)=(1,1)$ including the Manakov system ($n=2$)~\cite{vnls}.

Now if we introduce a new unified two-parameter $(\epsilon_x, \epsilon_t)$ symmetric reduction
\bee\label{con}
 \bR(x,t)=-\bQ^{\dag}(\epsilon_xx,\epsilon_tt),
\ene
with $(\epsilon_x, \epsilon_t)=(-1,1)$, then system (\ref{vnls}) with Eq.~(\ref{con}) is just one type of the above-introduced new nonlocal equation (\ref{vnlsp}). Therefore, Lax pair of Eq.~(\ref{vnlsp}) is given by Eqs.~(\ref{lax1}) and (\ref{lax2}) with new symmetric constraint (\ref{con}).

We introduce $n$ new complex functions~\cite{wadati} $\omega_j(x,t)=\frac{\psi_j(x,t)}{\psi_{n+1}(x,t)}\, (j=1,2,...,n)$ in terms of $n+1$ eigenfunctions $\psi_j(x,t)$ of Eqs.~(\ref{lax1}) and (\ref{lax2}) such that we find that $\omega_j(x,t)$ satisfy  $n$-component Riccati equations
\vspace{-0.03in}\bee
 \omega_{j,x}(x,t)\!=\!\sigma\omega_j(x,t)\!\sum_{s=1}^nq_s^{*}(\epsilon_xx,\epsilon_tt)\omega_s(x,t)
  -2i\lambda \omega_j(x,t)\!+\!q_j(x,t),\qquad\qquad
\label{ri}\ene
To solve Eq.~(\ref{ri}) we here consider their asymptotic (in $\lambda$) solutions and assume
their candidate solutions as power series expansions in parameter $2i\lambda$, with unknown functions of space and time as coefficients
 in the form
\bee\label{omega}
\begin{array}{l} \omega_j(x,t)=\sum_{s=0}^{\infty}\omega_{j}^{(s)}(x,t)(2i\lambda)^{-s-1}.
\end{array}
\ene
Substituting it into Eq.~(\ref{ri}) and comparing coefficients of terms $(2i\lambda)^{s}\, (s=0,1,2,...)$ to find
\bee
 \omega_{j}^{(0)}(x,t)=q_j(x,t), \,\,\, \omega_{j}^{(1)}(x,t)=-q_{j,x}(x,t),\qquad\qquad \\
  \omega_{j}^{(s+1)}(x,t)\!=\! \sigma\!\!\sum_{i=1}^nq_i^{*}(\epsilon_xx,\epsilon_tt)\!\!\sum_{k=1}^{s-1}\omega_{j}^{(k)}(x,t)\omega_{i}^{(s-k)}(x,t)
 -\omega_{j,x}^{(s)}(x,t), \,\, (s=2,3,...) \qquad\qquad
\ene

It follows from Eqs.~(\ref{lax1}) and (\ref{lax2}) with condition (\ref{con}) that
$
 (\ln |\psi_{n+1}|)_x\!=\!i\lambda-\sigma F(x,t), \,
 (\ln |\psi_{n+1}|)_t\!=\!2i\lambda^2\!-\!\sigma G(x,t)$,
whose compatibility condition, $(\ln |\psi_{n+1}|)_{xt}=(\ln |\psi_{n+1}|)_{tx}$ yields
\bee\label{conver} F_t(x,t)=G_x(x,t),
\ene
where $G(x,t)=\sum_{j=1}^n[2\lambda q_j^{*}(\epsilon_xx,\epsilon_tt)+iq_{j,x}^{*}(\epsilon_xx,\epsilon_tt)]\omega_j(x,t)
   +i\sum_{j=1}^nq_j(x,t)q_j^{*}(\epsilon_xx,\epsilon_tt)$ and $F(x,t)=\sum_{j=1}^n q_j^{*}(\epsilon_xx,\epsilon_tt)\omega_j(x,t)$.

Substituting Eq.~(\ref{omega}) into the conversed Eq.~(\ref{conver}) and comparing the coefficients of same terms $\lambda^j$ yields the infinite number of conservation laws. For example,
\bee
\partial_t\!\!\!\sum_{j=1,2}\!\!\!q_j(x,t)q_j^{*}(\epsilon_xx,\epsilon_tt)\!=\!i\partial_x\!\!\sum_{j=1,2}\!\![q_j(x,t)q_{jx}^{*}
(\epsilon_xx,\epsilon_tt)
+q_{jx}(x,t)q_j^{*}(\epsilon_xx,\epsilon_tt)]. \qquad\qquad
 \ene
Thus $\mathcal{Q}_{\epsilon_x,\epsilon_t}^{(n)}$ model (\ref{vnlsp}) is an integrable system for any parameter choice $(\epsilon_x, \epsilon_t)=(\pm 1, 1)$. Notice that the inverse scattering method~\cite{vnlsb} can also be extended to solve Eq.~(\ref{vnlsp}) by means of its Lax pair (\ref{lax1}) and (\ref{lax2}) with (\ref{con}), which will be given in another literature because of the page limit. Notice that the integrability of $\mathcal{Q}_{\epsilon_x,\epsilon_t}^{(n)}$ model (\ref{vnlsp}) with $(\epsilon_x, \epsilon_t)=(\pm 1, -1)$ is  not known.

{\it One-to-one connection between  two-parameter family and $\bPT$ symmetry and applications in $\mathcal{Q}_{\epsilon_x,\epsilon_t}^{(n)}$ model.}---We know that two-parameter $(\epsilon_x, \epsilon_t)$ family  determines four distinct $\mathcal{Q}_{\epsilon_x,\epsilon_t}^{(n)}$ models. In the following we will show that two-parameter $(\epsilon_x, \epsilon_t)$ family is also subtly related to $\bPT$ symmetry.  In the Introduction, we have  simplified $\bP$ abd $\bT$ symmetries as $\mathcal{P}:$\, $(x, t,i)\rightarrow (-x, t, i)$ and $\mathcal{T}:$\, $(x,t,i)\rightarrow (x, -t, -i)$ in terms of three variables $(x,\, t,\, i)$. Since mappings for $t$ and $i$ are always same in both $\mathcal{P}$ and $\bT$ (i.e., $\bP:(t,i)\rightarrow (t,i)$ and $\bT:(t,i)\rightarrow (-t,-i)=-(t, i)$). Thus we can introduce a  `new' variable $\tau=(t, i)$ such that $\mathcal{P}$ and $\bT$ can be further simplified as $\mathcal{P}:$\, $x\rightarrow -x$,\, ($\tau\rightarrow \tau)$ and $\mathcal{T}:$\, ($x\rightarrow x$),\, $\tau \rightarrow -\tau$ using two variables $(x,\, \tau)$.

\begin{figure}[!ht]
	\begin{center}
\hspace{-0.1in}{\scalebox{0.35}[0.35]{\includegraphics{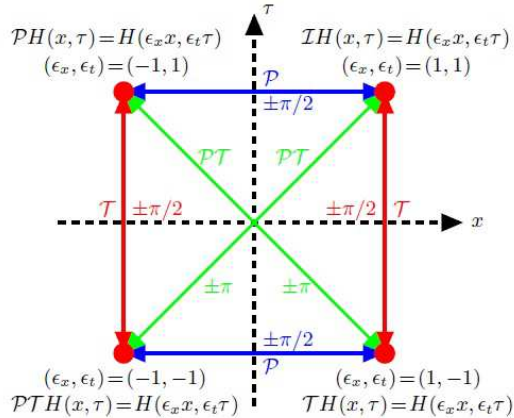}}}
	\end{center}
	\vspace{-0.25in} \caption{\small (color online)  The one-to-one  connection between symmetrility among four points $(\epsilon_x, \epsilon_t)=(\pm 1, \pm 1)$ in two-dimensional space $(x, \tau)$ and $\bPT$ symmetry with $H(\epsilon_xx, \epsilon_t\tau)\equiv H(\epsilon_xx, \epsilon_tt, \epsilon_ti)$.
$\pm \pi/2$ denote $90^{\circ}$ counterclockwise ($-$) and clockwise ($+$) rotations, respectively.} \label{PT}
\end{figure}

If the plane $\{(x, \tau)\}$ is regarded as a complex plane with $x, \tau$ being real and imaginary axes, respectively,
then four points $(\epsilon_x, \epsilon_t)\, (\epsilon_{x,t}=\pm1)$ correspond to complex numbers
$z_1=1+i,\, z_2=-1+i,\, z_3=-1-i,\, z_4=1-i$ such that $z_2=z_1e^{i\pi/2},\, z_3=z_2e^{i\pi/2},\, z_4=z_3e^{i\pi/2},\, z_1=z_4e^{i\pi/2},\, z_3=z_1e^{i\pi},\, z_4=z_2e^{i\pi}$, that is to say, every number can be obtained form another number by $m\pi/2$ counterclockwise
($m\in \mathbb{Z}^+$) or clockwise ($m\in\mathbb{Z}^-$) rotation (see Fig.~\ref{PT}), where $\mathbb{Z}^{+}$ ($\mathbb{Z}^{-}$) denotes the set of positive (negative) integers. Therefore we can perfectly establish a one-to-one connection from symmetries (rotations) among four points $\{(\epsilon_x, \epsilon_t)| \epsilon_{x,t}=\pm 1\}$ in two-dimensional space $(x, \tau)$ to $\{I,\,\bP,\, \bT,\, \bPT\}$ symmetries of the operator $H(x,\tau)$, where $\mathcal{I}$ is an identical operator (map) (see Fig.~\ref{PT}). The connection is new and simple to understand $\bPT$ symmetry from the plane figure.

For the $\bPT$-symmetric invariance of $(1+1)$-dimensional multi-component equations with $n$ distinct complex  fields $\phi_j(x,t)\, (j=1,2,\cdots, n)$ we have
\bee
 \bPT:\, \phi_j(x,t)\rightarrow \phi^{*}_{n+1-j}(-x,-t),
 \ene
where $j=1,\,2,..., (n+1)/2$ for $n=2\mathbb{Z}^{+}\!-\!1$ and  $j=1,\,2,..., n/2$ for $n=2\mathbb{Z}^{+}$. Notice that the above-mentioned $\bPT$-symmetric operator are also extended to the $(n+1)$-dimensional case where $x$ is replaced by ${\bf x}=(x_1,x_2,...,x_n)$.

In the following we show the $\bPT$ symmetrility of $\mQ(\epsilon_xx,\epsilon_tt)$ model.  For system (\ref{vnlsp}) with $n=1$, acting the $\bPT$-symmetric operator on the both side of Eq.~(\ref{nls1}) yields
 \bee
 iq^{*}_{1,t}(-x,-t)=-q^{*}_{1,xx}(-x,-t)+2\sigma q_1^{*2}(-x,-t)q_1(-\epsilon_xx,-\epsilon_tt),
\ene
which is equivalent to Eq.~(\ref{nls1}) since we know the $\bPT$ relation $q_1(x,t)=q^{*}_1(-x,-t)$ and the deduced relation $q_1(-\epsilon_xx,-\epsilon_xt)=q^{*}_1(\epsilon_xx,\epsilon_tt)$. This implies that Eq.~(\ref{nls1}) is $\bPT$ symmetric for any parameter choice. which is different from the invariance of the right-side operator of Eq.~(\ref{nls1}) with self-induced potential $-\partial_x^2+2\sigma q_1(x,t)q^{*}_1(-x,t)$ under the usual sense of $\bPT$ symmetry $\{x\to -x, i\to -i\}$~\cite{pt}.
Similarly, we can also show that new integrable local and nonlocal $\mQ_{\epsilon_x,\epsilon_t}^{(n)}$ model (\ref{vnlsp}) is $\bPT$ symmetric for any integer $n$ and two-parameter choice $(\epsilon_x, \epsilon_t)=(\pm 1, \pm 1)$.

Eq.~(\ref{vnlsp}) can be rewritten as $i\bQ_t(x,t)=\hat{H}_n(\epsilon_xx,\epsilon_tt)\bQ(x,t)$, where the Hamiltonian operator $\hat{H}_n$ with self-induced potential~\cite{self} is of the form
\bee \label{hn}
  \hat{H}_n(\epsilon_xx,\epsilon_tt)=-\partial_x^2+2\sigma \bQ^{\dag}(\epsilon_xx,\epsilon_tt)\bQ(x,t),
\ene
Now we consider the $\bPT$ symmetribility of Hamiltonian operators $\hat{H}_n$ using relations in Fig.~\ref{PT} (see Table~\ref{hnpt}). This indicates that only operator $\hat{H}_n(\epsilon_xx,\epsilon_tt)$ is $\bT$ symmetric for $(\epsilon_x, \epsilon_t)=(1,-1)$ and only operator $\hat{H}_n(\epsilon_xx,\epsilon_tt)$ is $\bPT$ symmetric for $(\epsilon_x, \epsilon_t)=(-1,-1)$. Notice that these terms including `No' in Table~\ref{hnpt} do not imply that those symmetries must not hold. They may admit some symmetries for some special potential $\bQ(x,t)$.
\begin{table}
\vspace{-0.1in}
\caption{\small $\bPT$ symmetribility of the Hamiltonian operators with self-induced potentials $\hat{H}_n$ given by Eq.~(\ref{hn}) for different two-parameter choices.\vspace{0.05in}}
 \begin{tabular}{ccccc} \hline\hline \\ [-3.0ex]
 \noindent  Case  &  $(\epsilon_x,\, \epsilon_t)$ \quad & \quad\,\, $\bP$ \,\,\quad & \,\,\quad $\bT$ \,\,\quad & \,\,\quad $\bPT$ \qquad  \\ \hline \\ [-3.0ex]
 i &  $(+1, +1)$  &  No & No &  No \\  \hline \\ [-3.0ex]
 ii &  $(-1, +1)$  &  No & No &  No \\  \hline \\ [-3.0ex]
 iii &  $(+1, -1)$  &  No & Yes &  No \\  \hline \\ [-3.0ex]
 iv &  $(-1, -1)$  &  No & No &  Yes \\  \hline\hline
\end{tabular}
\label{hnpt}
\end{table}

{\it Multi-linear and self-similar reductions.}---For the given equation (\ref{vnlsp}), if we can seek for some self-similar transformations reducing it to ones solved easily, then its solutions may be found. Here we have two special reductions, i) if $q_j(\epsilon_xx,\epsilon_tt)=q_j(x,t)$, then Eq.~(\ref{vnlsp}) reduce to Eq.~(\ref{nls0})  whose solutions are known; ii) if $q_j(\epsilon_xx,\epsilon_tt)=-q_j(x,t)$, then Eq.~(\ref{vnlsp}) reduce to Eq.~(\ref{nls0}) with $\sigma$ replaced by $-\sigma$ whose solutions are known. For example, if the solutions of Eq.~(\ref{nls0}) are even functions for space, then solutions of local Eq.~(\ref{nls0}) are also ones of Eq.~(\ref{nls1}). Now we consider rational transformations of Eq.~(\ref{vnlsp})
\bee \label{bi1}
 q_j(x,t)=\frac{g_j(x,t)}{f(x,t)}, \quad f(x,t),\, g_j(x,t)\in\mathbb{C}[x,t]
\ene
where $f(x,t),\, g_j(x,t)\in\mathbb{C}[x,t]$, which differ from the usual ones~\cite{hirota} such that we have multi-linear equations
\bee
\begin{array}{l} {\rm Bilinear \,\, eq.:}\,\,\, (iD_t+D_x^2-\mu) g_j(x,t)\cdot f(x,t)=0,\vspace{0.1in} \\
{\rm Trilinear \,\, eq.:}\,\, f^{*}(\epsilon_xx,\epsilon_tt)(D_x^2-\mu)f(x,t)\cdot f(x,t)
   =-2\sigma f(x,t){\bf G}^{\dag}(\epsilon_xx,\epsilon_tt){\bf G}(x,t), \end{array}
\label{bi2}\ene
where $\mu\in \mathbb{C}$, ${\bf G}(x,t)=(g_1(x,t), g_2(x,t),...,g_n(x,t))^T$ and ${\bf G}^{\dag}(\epsilon_xx,\epsilon_tt)$ is the transpose conjugate of ${\bf G}(\epsilon_x x,\epsilon_tt)$, $D_t$ and $D_x$ are both Hirota's bilinear operators defined by~\cite{hirota} $D_t^mD_x^n f\cdot g=(\partial_t-\partial_{t'})^m(\partial_x-\partial_{x'})^n[f(x,t)g(x,t)]|_{x=x',t=t'}$. For the case $(\epsilon_x,\epsilon_t)=(1,1)$, we know that
${\bf G}^{\dag}(\epsilon_xx,\epsilon_tt){\bf G}(x,t)$ is real-valuable function such that we can assume $f(x,t)\in\mathbb{R}[x,t]$ from
Eq.~(\ref{bi2}), which leads to $f^{*}(\epsilon_xx,\epsilon_tt)=f(x,t)$, in which Eq.~(\ref{bi2}) becomes a bilinear equation~\cite{hirota}.  Multi-wave solutions of Eq.~(\ref{vnlsp}) can be found in terms of Eqs.~(\ref{bi1})-(\ref{bi2}) and series expansions of  $g_j(x,t)$ and $f(x,t)$. Here we give one-soliton solution of Eq.~(\ref{vnlsp}) with $n=1$
\vspace{-0.03in}\bee
\begin{array}{l}
 q_1(x,t)=\dfrac{(\epsilon_xk^{*}+k)^2e^{kx+ik^2t}}{(\epsilon_xk^{*}+k)^2-\sigma e^{(\epsilon_xk^{*}+k)x+i(k^2-\epsilon_tk^{*2})t}}, \quad
\end{array}\ene
where $k\in\mathbb{C},\, \epsilon_xk^{*}+k\not=0$, and $k^{*}$ is a complex conjugate of $k$.

Now we apply the direct reduction method~\cite{ck} to consider the self-similar solution of Eq.~(\ref{vnlsp})
\vspace{-0.03in}\bee \label{si1a}
\begin{array}{l}
q_j(x,t)=\frac{p_1(z)}{\sqrt{2t}}e^{i\mu_j\log|t|/2}, \,\,\, z(x,t)=x/\sqrt{2t},
\end{array}\ene
where $\mu_j\in\mathbb{R}$, $p_j(z)\in \mathbb{C}[z],\, (j=1,2,...,n)$, and $ x,t\in \mathbb{R}$, which differs from one used in~\cite{am} where $t>0$ is required. Thus we have the equation satisfied by $p_j(z)$
\vspace{-0.03in} \bee \label{si1b}
  p_{j,zz}\!-\!(i+\mu_j)p_j(z)\!-\!izp_{j,z}(z)
  -\!2\sigma\sqrt{\epsilon_t}p_j(z)\sum_{j=1}^np_j(z)p_j^{*}(\hat{z})\!=\!0,
\ene
via the substitution of Eq.~(\ref{si1a}) into Eq.~(\ref{vnlsp}), where $\hat{z}=\epsilon_x x/\sqrt{2\epsilon_tt}$. Notice that i) when $(\epsilon_x, \epsilon_t)=(1,1)$, we have $\hat{z}=z$ and $p_j^{*}(\hat{z})=p_j^{*}(z)$ such that the self-similar reduction becomes the usual result; ii) when $(\epsilon_x, \epsilon_t)=(-1,1)$, we have $\hat{z}=-z$ and $p_j^{*}(\hat{z})=p_j^{*}(-z)$ such that the self-similar reduction given by Eq.~(\ref{si1a}) and (\ref{si1b}) with $j=1,\, t>0$ becomes the result~\cite{am} (notice that $i$ should be $-i$ in Eq.~(30) of Ref.~\cite{am}); iii) when $(\epsilon_x, \epsilon_t)=(1,-1),\, (-1,-1)$, the self-similar reduction  of Eq.~(\ref{vnlsp}) given by Eqs.~(\ref{si1a}) and (\ref{si1b}) are new.

Similarly, we consider another self-similar solution of separating variables
$q_j(x,t)=p_j(x)e^{\omega_j\sqrt{-\epsilon_t}t},\, (j=1,2,...,n)$
where $\omega_j\in \mathbb{R}$,\, $p_j(x)\in \mathbb{C}[x]$, and $ x,t\in \mathbb{R}$. The substitution of this transformation into Eq.~(\ref{vnlsp}) yields
 \vspace{-0.05in}\bee\label{si2b}
   p_{j,xx}(x)-2\sigma p_j^2(x)p_j^{*}(\epsilon_xx)+i\omega_j \sqrt{-\epsilon_t} p_j(x)=0.\quad
\ene
When $(\epsilon_x, \epsilon_t)=(-1,1),\, (1,1)$, we can assume $p_j (x)\in \mathbb{R}[x]$, in which our result reduces to the know one~\cite{am}.
Whereas $(\epsilon_x, \epsilon_t)=(1,-1),\, (-1,-1)$, we may not assume $p_j (x)\in \mathbb{R}[x]$ from Eq.~(\ref{si2b}) and it should be a complex function. Other similar solutions of Eq.~(\ref{vnlsp}) may be found using Lie classical and non-classical symmetric methods~\cite{sm}.

 {\it Exact soliton solutions.}---Some {\it singular} (breather) solutions of focusing ($\sigma<0$) Eq.~(\ref{nls1}) had been found using the inverse scattering method~\cite{am}. In the following we present its some {\it regular} soliton solutions.

i) {\it double-periodic and soliton solutions:} Similarly, we have the double-periodic wave solutions of Eq.~(\ref{nls1}) with the focusing interaction $\sigma<0$
\vspace{-0.03in}\bee\begin{array}{l}
 q_{1cn}(x,t)=mk/\sqrt{-\sigma}\,{\rm cn}(kx, m)e^{i (2m^2-1)k^2t},  \vspace{0.05in}\\
 q_{1dn}(x,t)=k/\sqrt{-\sigma}\,{\rm dn}(kx, m)e^{i (2-m^2)k^2t},  \vspace{0.05in}\\
  q_{1sn}(x,t)=mk/\sqrt{-\sigma}\,{\rm sn}(kx, m)e^{-i (1+m^2)k^2t}, \vspace{0.05in}\\
 q_{1scd}(x,t)=\frac{a m k {\rm cn}(kx, m)+M{\rm sn}(kx,m)}
 {mk+\sqrt{m^2k^2+4\sigma a^2}{\rm dn}(kx, m)}e^{i (m^2-2)k^2/2 t},
\end{array}\ene
where $M=imk/2\sqrt{4a^2(1-m^2)-m^4k^2/\sigma}$, $m\in (0, 1)$ is the modulus of Jacobi elliptic functions, $k,\, a$ are real parameters. In particular, for $m\rightarrow 1$, we know that $ q_{1cn}(x,t)$ and $q_{1dn}(x,t)$ both reduce to the bright soliton $q_{1b}(x,t)=k/\sqrt{-\sigma}\,{\rm sech}(kx)e^{i k^2 t}$~\cite{li}, $q_{1sn}(x,t)$  reduces to the dark soliton $q_{1d}(x,t)=k/\sqrt{-\sigma}\/\tanh(kx)e^{-2i k^2 t}$~\cite{li}, and
$q_{1scd}(x,t)$  reduces to the new combination solution of $q_{1d}(x,t)$ and $q_{1b}(x,t)$
\bee
 \begin{array}{l} q_{1bd}(x,t)\!=\!\dfrac{2ak{\rm sech}(kx)\!+\!ik^2/\sqrt{-\sigma}\,\tanh(kx)}{2k\!+\!2\sqrt{k^2+4\sigma a^2}{\rm sech}(kx)}e^{-\frac{i k^2 t}{2}}, \qquad
\end{array}\ene
where $a, k$ are constants.

  We have $q_{1b}(x,t)=q_{1b}^{*}(-x, -t)$, but $q_{1b}(x,t)\not\equiv q_{1b}^{*}(-x, t)$ except for the trivial case $k=0$. Thus the bright soliton is generalized $\bPT$-symmetric.  We know $q_{1d}(x,t)=-q_{1d}^{*}(-x, -t)$, but $q_{1d}(x,t)\not\equiv q_{1d}^{*}(-x, t)$ except for the trivial case $k=0$. Thus the dark soliton is not $\bPT$-symmetric. For $a\not=0,\, k>2\sqrt{-\sigma a^2}$ and $a=0$, $q_{1bd}(x,t)$ is a dark soliton, but for $a\not=0,\, k<-\sqrt{k^2-4\sigma a^2}$, $q_{1bd}(x,t)$ is a bright soliton. For $a\not=0$, $q_{1bd}$ is an neither even nor odd function for space $x$. For $a=0$, $q_{1bd}$ is an odd function for space $x$. We have $q_{1bd}(x,t)=q_{1bd}^{*}(-x, -t)$, but $q_{1bd}(x,t)\not\equiv q_{1bd}^{*}(-x, t)$ except for the trivial case $k=0$. These double-periodic wave solutions also have similar $\bPT$-symmetric properties.

ii) {\it multi-rogue wave solutions:} We find that the multi-rogue wave solutions of the focusing NLS equation are both even for space (or under some transformations)~\cite{nail}. Thus we have multi-rogue wave solutions of Eq.~(\ref{nls1}) with $\sigma<0$ in terms of ones of the focusing NLS equation. For example, the first-order rogue wave (rogon) solution of Eq.~(\ref{nls1})  with $\sigma<0$ is
$q_{1rw}(x,t)=\left[1-\frac{4(1+4it)}{1+4x^2+16t^2}\right]\!e^{2it}$.

{\it Conclusion and discussion.}---We have first introduced a unified two-parameter model from a new and simple two-parameter symmetric reduction of vector NLS system. The two-parameter model just connects one local and three nonlocal vector NLS equations. It is shown to possess a Lax pair and infinite number of conservation laws and to be $\bPT$ symmetric. We also give its
multi-linear form and some self-similar solutions as well as some explicitly exact regular solutions including bright and dark solitons, double-periodic wave solutions and multi-rogue wave solutions.  Moreover, we also establish a one-to-one connection between a one-to-one
connections between symmetribility of four points $\{(\epsilon_x, \epsilon_t)|\epsilon_{x,t}=\pm1\}$ and $\{{\mathcal I}, \bP, \bT, \bPT\}$ symmetries.
The $\mQ_{\epsilon_x,\epsilon_t}^{(n)}$ model related to vector NLS equations will provide more novel integrable models in integrable systems. In fact, the used two-parameter family can also be extended to the higher-dimensional~\cite{hd} and discrete systems~\cite{vnlsb,amd}. For example, a unified new two-parameter discrete $\bPT$-symmetric vector model is generated
\bee \begin{array}{l} i{\bf P}_{n,t}(t)=-{\bf P}_{m+1}(t)+2{\bf P}_{m}(t)-{\bf P}_{m-1}(t)
+\sigma\sum_{j=1,2}{\bf P}_{m+2-j}(t){\bf P}_{\epsilon_x m}^{\dag}(\epsilon_tt){\bf P}_{m+1-j}(t) \end{array} \ene
 with $m\in \mathbb{Z}$,\, ${\bf P}_m(t)=(p_{1,m}(t),p_{2,m}(t), ...,
p_{n,m}(t))^T$ and ${\bf P}_{\epsilon_x m}^{\dag}(\epsilon_tt)$ being the transpose conjugate of ${\bf P}_{\epsilon_x m}(\epsilon_tt)$, which is regarded as an integrable discretization of Eq.~(\ref{vnlsp}).

We know that the introduced two-parameter family $(\epsilon_x,\epsilon_t)$ plays a central role in the study of both new integrable models and $\bPT$ symmetry. In fact, we may consider more general two-parameter family as $(\epsilon_x, \epsilon_t)$ with $\epsilon_{x,t}=\pm 1, \pm i,\ (i=\sqrt{-1})$, thus we have sixteen possible two-parameter choices, i.e., $(\epsilon_x, \epsilon_t)\in \{(1,1)$, $(-1, 1),$ $(-1, -1)$, $(-1,1)$, $(i,1)$, $(-i, 1)$, $(-i, -1)$, $(i, -1)$, $(1, i)$, $(-1, i)$, $(-1, -i)$, $(1, -i)$, $(i, i)$, $(-i, i)$, $(-i, -i)$, $(i, -i)\}$, in which the first four components have been considered in Eq.~(\ref{vnlsp}). For other twelve families of two-parameter choices, the $\bPT$ symmetry may not be powerful and will be enlarged and the $\mQ_{\epsilon_x,\epsilon_t}^{(n)}$ model may exhibit different integrability and wave structures. These will be studied in another literature.

\vspace{0.1in} \noindent {\bf Acknowledgments}

 The author would like to thank for the referees for their valuable suggestions.  This work was partially supported by the NSFC under Grant No.61178091 and  the NKBRPC under Grant No. 2011CB302400.




\end{document}